\documentclass[aps,twocolumn,prl,color,psfig,epsf,showpacs]{revtex4}
\usepackage{color}
\usepackage{amsmath}
\usepackage{amsfonts}
\usepackage{times}
\usepackage{epsf}
\usepackage{graphicx}

\def\thetao{{\theta_0}}
\begin{document}

\title{Stall, spiculate or runaway - the fate of fibers growing towards fluctuating membranes}

\author{D. R. Daniels$^1$, D. Marenduzzo$^2$, M. S. Turner$^1$}

\affiliation{$^1$ Department of Physics, University of Warwick,
Coventry CV4 7AL, UK\\
$^2$ SUPA, School of Physics, University of Edinburgh, Mayfield Road, 
Edinburgh EH9 3JZ, UK}

\begin{abstract}
We solve the dynamic equations of motion for a growing semi-flexible 
polymer, or fiber, approaching a fluctuating membrane at an angle.
At late times we find three different regimes: fiber {\em stalling},
when fiber growth stops due to membrane resistence,
{\em run-away}, in which the polymer bends away from the membrane, 
and another regime in which the membrane response is nonlinear and tubular 
membrane {\em spicules} are formed. We discuss which regions of the
resulting `phase diagram' are explored by (i) single and bundled actin 
fibers in living cells, (ii) sickle hemoglobin fibers in red blood 
cells, and (iii) microtubules growing within artificial vesicles. 
We complement our analysis with full 3-dimensional 
stochastic simulations. 
\end{abstract}

\maketitle

Semi-flexible polymer fibers, making up the cytoskeleton, interact with 
fluctating membranes continuously in living cells \cite{bray}. The 
polymerisation of filaments such as actin, with persistence lengths 
$l_p\sim 10\mu$m, can deform the outer plasma membrane of the cell, giving 
rise to protrusions that can be sheet-like (lamellipodia) or more localised 
(filopodia) \cite{bray}. Forces are thought to arise through polymerisation 
at the fiber tips \cite{oster,mogilner}.  A similar mechanism is thought to 
drive growth of sickle hemoglobin fibers, leading to a pathological  
rigidification of red blood cells \cite{eaton}, as well as microtubules, such 
as can be grown in artificial vesicles \cite{phi}.

While the phenomenon of Euler buckling of a fiber of fixed length impinging 
on a solid obstacle at an angle is well studied (see e.g. \cite{dickinson} 
and Refs. therein), the case in which the elastic fiber is growing, and the 
obstacle is  a fluctuating membrane, have so far received much less attention, 
despite its more direct relevance to biological systems. Thus
here we propose a set of equations of motion that describe the coupled 
dynamics of a growing  semi-flexible polymer close to a fluctuating membrane 
(Fig~1). 
{ {We complement this treatment with full
3-dimensional dynamic Monte-Carlo (3dMC) 
simulations. 
Our main focus is on the dynamic regimes attained at
late times. We find that fibers may (i) {\em stall},
(ii) {\em run-away}, i.e. bend away from the membrane, 
or (iii) cause the formation of tubular
membrane {\em spicules}.
What differentiates our approach from earlier work 
is the explicit description of the flexibility of both the 
membrane and the polymer 
\cite{mogilner,dickinson,mogilner_filopodia,nigel,derenyi},
which allows us to include all three fates of the
fiber.}} 

The Hamiltonians governing the membrane and the polymer elasticity,
respectively $H_m$ and $H_p$, are given by :
\begin{eqnarray}\label{hamiltonian}
H_m & = & \frac{1}{2} \int d^2 r \, 
\Big( \kappa_{m} ( {\nabla}_{\perp}^2  u )^2 + 
\sigma ( \nabla_{\perp}  u )^2 \Big) \\ \nonumber
H_p & = & \frac{\kappa}{2} \int_{0}^{L} ds \, 
\Big( \frac{\partial \theta}{\partial s} \Big)^2 
\end{eqnarray}
where $\kappa_m$ and $\sigma$ denote the membrane rigidity and
surface tension respectively and are typically  $\kappa_m\sim 10^{-19}$ J 
and $\sigma\sim 10^{-4}$ Jm$^{-2}$ for biomembranes, $u$ measures the 
normal deviation of the membrane from local flatness, which we take to be the 
plane $z=0$, $\kappa=k_BT l_p\equiv l_p/\beta$ 
is the polymer bending rigidity, $L$ is its 
(instantaneous) length and $\theta(s,t)$ is the angle between the local 
direction of the fiber at arc length position $s$ and the plane 
$z=0$ at time $t$, see Fig~1. 
Our use of an explicit Hamiltonian to describe the flexibility of both the 
membrane and polymer differentiates our approach from earlier work 
\cite{mogilner,dickinson,mogilner_filopodia,nigel,derenyi}. 
The probability distribution 
of the membrane displacement can be calculated 
from the conditional partition function that it is at fixed height $z_m$ 
at one point. This is here approximated by \cite{rob} 
${\cal{Z}}_m = \exp (- A  z_m^2 )$,
where $A = \frac{2 \pi \sigma}{\log \big( 1+ \frac{\sigma \Omega}
{\kappa_{m}\pi^2} \big)}$, 
$\Omega$ being the
area of the ``frame" supporting the membrane patch under consideration.
{ {{$A$} measures the membrane 'softness'
($A\to\infty$ corresponds to a hard wall), and 
typically $A\sim 10^{-4}$ Jm$^{-2}$~\cite{rob}.}}
The partition function ${\cal{Z}}_m$ can be integrated over all allowed $z_m$ 
to obtain the  normal force acting on the fiber tip
due simply to the steric repulsion between tip and membrane. As is common in 
studies of buckling we perform a saddle point approximation on $H_p$ in 
Eq. \ref{hamiltonian}, and consider
the adiabatic limit in which, at all times, the fiber is in 
equilibrium. This 
is reasonable in all physical cases as the fiber relaxation time (at most ms) 
is much smaller than the inverse polymerisation rate (typically 0.01-1 s). 
We seek to solve the following coupled equations for the fiber shape, 
growth rate and tip height and the normal force on the tip, respectively
\begin{eqnarray}
0 &=&\left[ \kappa\frac{\partial^2 \theta(s,t)}{\partial s^2}
-f\cos\left(\theta(s,t)\right)\right] \label{saddle_pol}\\
\frac{\partial L(t)}{\partial t} & = & \delta\left\{k_{\rm on}
e^{\left[-\beta f(\Delta)\delta
\sin\left(\theta(L(t),t)\right)\right]}-k_{\rm off}\right\}  \label{eq2} \\
\Delta(t)& = & -d+\int_0^{L(t)} ds \sin\left(\theta(s,t)\right) \label {eq3}\\
f(\Delta)  &=& 2\sqrt{\frac{A}{\pi}}
\frac{\exp\left(-A\left(\Delta(t)\right)^2\right)}
{1-{\rm erf}\left(\sqrt{A}\Delta(t)\right)} \label{force}
\end{eqnarray}
Here $k_{\rm on,off}$ are, respectively, the rates of polymerisation and 
depolymerisation at the fiber tip, $d$ and $\Delta$ 
the initial and instantaneous distances between the fiber tip 
and the membrane frame, $\delta$ the increase in length upon
addition of a monomer, while $f(\Delta)$ is the
force exerted by the membrane on the fiber tip \cite{rob}. 
We assume constant rates $k_{\rm on,off}$ but time varying rates 
could be included to model, e.g. actin concentration fluctuations. 
The fiber is treated as if clamped at one end $\theta(s=0,t)\equiv
\theta(s=0)\equiv \thetao$, into the cytoskeletal mesh, while the end at 
$s=L(t)$ is free,
$\left({\partial \theta(s,t)}/{\partial s}\right)_{s=L(t)}=0$. 
Eq. \ref{force} has been derived by assuming that the membrane and the 
fiber only interact via excluded volume and that this can 
be introduced at a single point (the fiber tip) \cite{rob}. 
This may fail for a highly bent fiber that often contacts the membrane 
far from its tip \cite{note}.
We solve  Eqs. \ref{saddle_pol}-\ref{eq3} numerically, via a 
standard Euler relaxation algorithm and finite difference discretisation.
The solutions for the system of Eqs. \ref{saddle_pol}-\ref{force} fall into 
two classes. The fiber can either (a) grow until it stalls, or (b) undergo a 
run-away transition ~\cite{nigel}.
This is distinct from usual fiber buckling, which is a purely
equilibrium phenomenon. The physical parameters determining the system 
dynamics are the four dimensionless quantities ${\sqrt{\frac{A}{k_BT}}\delta}$,
${k_{\rm on}/k_{\rm off}}$, $\frac{\kappa\delta}{k_BTd^2}$, and $d/\delta$.
{{Eq. \ref{saddle_pol} neglects fluctuations around the
average fiber shape (see below for a discussion of the effect of these).
Throughout we work at physiological temperature $T$.} 

\begin{figure}
\centerline
{\includegraphics[width=4.cm]{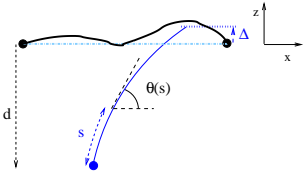}}
\caption{Sketch of the system under consideration. A growing fiber is 
obliquely incident onto a fluctuating membrane, which is anchored to some 
distant frame. The fiber is clamped at $s=0$ and its length at time $t$ is 
$L(t)$.
}
\end{figure}

Fig. 2 shows a cut of the `phase diagram': for $\thetao>\theta_c$ the fiber 
ultimately stalls, otherwise it is bent away by the membrane. Here $\theta_c$ 
is established by variation of $k_{\rm on}/k_{\rm off}$, for parameters 
typical of actin fibers in cells (see caption). 

\begin{figure}
\centerline
{\includegraphics[width=5.cm]{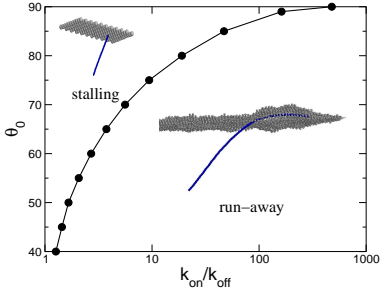}}
\caption{Plot of the critical angle $\theta_c$ (solid line) below which
a run-away transition occurs, as a function of $k_{\rm on}/k_{\rm off}$,
for actin-like parameters ($d=100$ nm, $\kappa=10$ $k_BT$ $\mu$m, 
$\sqrt{A}\delta\sim 0.8$, $\delta=2.5$ nm). 
(The run-away state shown on the right is
obtained with 3dMC simulations with $\thetao=45^\circ$, 
$k_{\rm on}/k_{\rm off}=\infty$ and 
$\sqrt{A}\delta=0.4$.)}
\end{figure}

As the incidence angle approaches the critical threshold
from below, the growth of the fiber is strongly 
non-linear at intermediate times (see Fig. 3a). This is because the 
growth rate is reduced by a factor exponential in 
$f(\Delta)$ (see Eq. (2)). This force bends the tip (Eq. (1)) and as a
result the growth speeds up again, as the factor $\sin(\theta(L(t),t))$
decreases (Eq. (2)). Far from the transition line
this behaviour is not found and the fiber growth is nearly 
linear at all times, see Fig. 3a.

\begin{figure}
\centerline
{\includegraphics[width=7.cm]{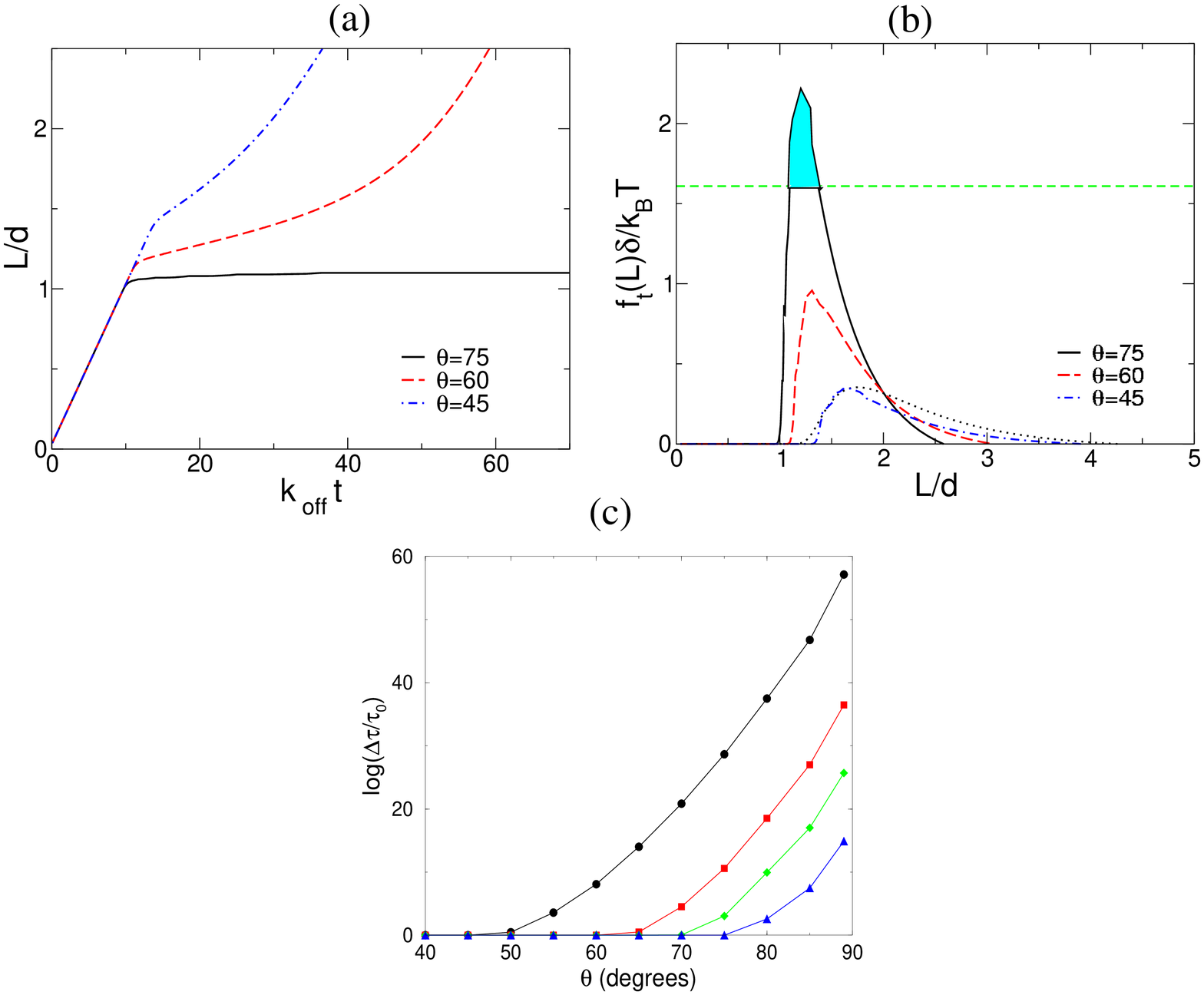}}
\caption{For three different initial incidence angle $\thetao$ we show (a) the 
fiber length as a function of time, and 
(b) the tangential adiabatic force $f_t$ as 
a function of length. Here $k_{\rm on}/k_{\rm off}=5$ (other parameters as 
Fig. 2), so that the stalling force is $\log{(5)} k_BT/\delta$, 
and the fiber is stalled for $\theta_0=75^{\circ}$ and runs away
at the two smaller incidence angles. This diagram serves to 
show that {\it all} stalled states are metastable. The analytical prediction 
for the adiabatic
longitudinal force (dotted line) is compared to the numerical result for
$\thetao=45^{\circ}$.
The time needed to tunnel the barrier  to 
run-away growth can be estimated as being proportional 
to the exponential of an energy barrier, being the area of the coloured 
region for $\thetao=75^{\circ}$ in (b).
The linear-logarithmic plot of this 
tunneling time as a function 
of its initial incidence angle, found by an application of Kramers' theory,
is shown in (c). From top to bottom, $k_{\rm on}/k_{\rm off}=1.1$,
$1.5$, $3$, $5$ and $10$.}
\end{figure}

The physics of the transition between the metastable stalled state and
the run-away state is controlled by the {\em tangential}
force on the fiber tip $f_{t}(L)$, see Fig. 3b -- $f_t$
is related to the normal force in Eq. \ref{force} via a factor 
$\sin(\theta(L))$ to be determined self consistently from Eqs. 
\ref{saddle_pol}-\ref{force}. If the thermodynamic stall force 
$f_{stall}=(k_BT/\delta)\log(k_{on}/k_{off})$ is above the maximum of $f_t$ 
the fiber runs away, otherwise the fiber 
stalls when it reaches the length at 
which the membrane generated force equals its thermodynamic stalling force. 
However, as the adiabatic force vanishes for very long fibers, the stalled
state is, in fact,  metastable and the fiber may still run away 
provided it can `tunnel' through the free energy barrier present. 
Using elementary transition state theory, we can estimate the time 
required for such a tunneling event as
$\Delta \tau = \tau_0 \exp \Big(\int_{L_1}^{L_2} dL \, f_t \Big)$
where $\tau_0\sim$ ms 
is a typical microscopic relaxation time (or inverse attempt 
frequency) and $L_1$ and 
$L_2$ are given by intercepts of the force curve with the 
thermodynamic stall force, see Fig. 3b. Fig. 3c shows a representative plot 
of $\Delta \tau$ versus 
$\theta_0$, with actin-like parameters. 

Actin fibers
in cells can be biochemically capped \cite{bray}. If the
tunneling time is larger than the inverse capping rate, typically
$< 1$ s, the fiber growth will be halted. Tunneling from the metastable 
stalled state may thus often be negligible .

While Eqs. \ref{saddle_pol}-\ref{force} have to be solved numerically, 
one can employ a Gaussian approximation for the fiber tip 
fluctuations, controlled by $H_p$ in (\ref{hamiltonian}), which permits some 
analytic analysis and, in particular, yields an explicit formula for $f_t(L)$. 

To achieve this, starting from Eq. \ref{hamiltonian}, we adopt a standard
path integral method  \cite{rob}, and compute the first two moments of the 
probability distribution for the fiber tip position
\begin{eqnarray}
\langle z_p \rangle_0 \,  & = & 
\, 2 \kappa \sin \theta_0 
\Big( 1 - \exp \big( - \frac{L}{2 \kappa} \big) \Big)
\label{zav} \\
\langle z_p^2 \rangle_0 \,  & = & 
\, 4 \kappa^2 \bigg( \frac{L}{2 \kappa} 
+ \exp \big( - \frac{L}{2 \kappa} \big) 
-1 - \frac{\cos 2 \theta_0}{12} \\ \nonumber
\, & & \, \Big( 3 + \exp \big( - \frac{2 L}
{\kappa} \big) - 4 \exp \big( - \frac{L}{2 \kappa} \big) \Big) \bigg)
\label{z2av}
\end{eqnarray}
where the subscript $0$ denotes the fact that these are derived in the 
absence of a membrane.
Using these moments we can approximate the distribution of
heights $z_p$ of the tip of the semiflexible polymer 
by a Gaussian distribution, 
${\cal{Z}}_p = \exp \Big( - B \, \big( z_p - \langle z_p \rangle_0 \big)^2 \Big)$
where $B = \frac{1}{2} \, \frac{1}{\langle z_p^2 \rangle_0 
\, - \, \langle z_p \rangle_0^2}$.
This approach is appropriate provided the force on the 
fiber is $\ll$ the  Euler buckling threshold, $\pi^2\kappa/(4L^2)$. So far 
we have calculated {\it separately} : (i) the membrane deformation, 
and (ii) the fiber tip distributions. 
We now introduce the steric {\it interaction} between the rod and membrane as 
follows :
\begin{eqnarray}
{\cal{Z}}_{tot} & = & \int_{-\infty}^{\infty} dz_m \int_{-\infty}^{\infty} 
dz_p \; {\cal{U}} ( z_m - z_p ) \;  {\cal{Z}}_m \; {\cal{Z}}_p \nonumber \\
& = & \frac{1}{2} \bigg( 1 - \mbox{erf} \, \Big({{\langle z_p \rangle_0 
- d }\over{\sqrt{1/A+ 1/B}}} \Big) \bigg)
\label{totpartition}
\end{eqnarray}
where ${\cal U}(x)=1$ if $x>0$ and $0$ otherwise.

It is reassuring that for completely rigid fibers we recover the corresponding 
result derived in Ref. \cite{rob}. Using ${\cal{Z}}_{tot}$ one can 
calculate, e.g. the tangential force acting on the tip
$f_{L} = - \frac{\partial \log ( {\cal{Z}}_{tot} )}{\partial L}$. 
This can be directly compared with the full numerical treatment
(see Fig. 3b, in which the comparison is shown for a fiber
with $\thetao=45^{\circ}$). 
For small incidence angle, up to $\thetao\approx 55^{\circ}$ (with
the parameter values of Figs. 2-3), the agreement is good, breaking down
gradually as the initial incidence angle approaches normal incidence.
{ {The smoother force rise in the analytics for
$\thetao=55^{\circ}$ is due to
the inclusion of tip fluctuations, which do not alter the value of the
maximum.}}
(We note that if the membrane is floppier than the agreement persists
to a larger angle.)
To see why the analytics eventually break down, 
we note that invoking a Gaussian tip 
distribution amounts to replacing the fiber with a spring with a Hookean 
constant $B$, which becomes
stiffer with $\thetao$. 
For $\theta$ not too close to $90^{\circ}$, 
$B\simeq \frac{3\kappa}
{2L^3\cos^2(\theta)}$, in agreement with the treatment of 
Ref. \cite{dickinson}. 
Our combined treatment of fiber tip and membrane
fluctuations using a Gaussian distribution can be seen as the natural 
generalisation of the elastic fiber model of Refs. 
\cite{mogilner,dickinson} to the case of a fluctuating soft membrane. 
As the incidence angle becomes close to normal, intuition suggests
that the tip fluctuations will become `one-sided' and the Gaussian
approximation may then become poor.

We note from Fig. 2 that in an eukaryotic cell 
(where $k_{\rm on}/k_{\rm off}\sim 100$~\cite{mogilner}),  
$\theta_c\sim 86^{\circ}$ for an actin fiber.
Thus the majority of fibers in a cytoskeletal actin network
may be expected to eventually run away from the membrane. Fibers stall at smaller angles 
than this if the initial fiber-membrane separation is smaller 
(than $d=100$ nm).
Electron microscopy of the actin network \cite{tomography} reveals that some 
of the fibers {\it do} appear to be bent into a sub-membrane `thatchwork', 
which may help the cytoskeleton sustain stress although might contribute less 
efficiently to motility. However in lamellipodia and filopodia,
observed at the leading edge of a moving cell, the fibres are actively 
pushing with their tips pointing towards the membrane \cite{filopodia}.
It is then natural to ask what `countermeasures' a moving 
cell takes to avoid run away.

One efficient strategy 
is to combine many fibers into a thicker, thus stiffer,
bundle. If $n$ fibers are cross-linked into one 
bundle, e.g. 
by the Ena/VASP proteins or the `tip complex' \cite{filopodia}, the aggregate 
can still be described as a single fiber obeying Eqs. 
\ref{saddle_pol}-\ref{force}, up to a rescaling
$\delta \rightarrow  \frac{\delta}{n}$,
$k_{on, off} \rightarrow n \, k_{on, off}$ and
$\kappa \rightarrow n^2 \, \kappa$, which amounts to assuming
that the fibers are strongly cross-linked.
As a result of bundling both stalling and run-away occur 
at larger membrane forces.

Fig. 4 shows how the dynamic `phase diagram' of Fig. 2 gets modified
by bundling. Here we have taken $n=10$,
which may be a reasonable assumption for small filopodia {\it in vivo}
\cite{mogilner_filopodia}. 
Solving Eqs. \ref{saddle_pol}-\ref{force} shows that $\theta_c$ 
decreases with $n$, so that the stalling regime widens (Fig. 4).
More importantly, a stiffer fiber, or a bundle with
larger $n$, will deform the membrane more.
As a result if $n$ is large enough, and $\thetao$ is close to $90^\circ$, non-linear membrane deformation into 
tubular spicules \cite{derenyi} may occur before run-away. 
We adopt a rough test for when spicules will be formed that 
is related to the 
breakdown of the 
linear membrane Hamiltonian (1), specifically that the maximum average membrane 
gradient exceeds unity $\langle |\nabla_{\perp} 
u |_{max} \rangle>1$. It can be shown \cite{rob} that this spiculation 
condition can be translated to a condition on the average membrane 
displacement, namely that it exceeds
$\langle z_m \rangle = d + \frac{1}{2A} 
\frac{\partial \log ( {\cal{Z}}_{tot} )}{\partial d}$, which can be calculated
from (\ref{totpartition}) 
$\langle z_m \rangle - d \; = \;  \frac{5 \pi}{2 A} 
\, \sqrt{\kappa_{m} \sigma}$. 
Inserting appropriate physical values we estimate that spicules form only 
if the (maximum) membrane displacement is at least $\sim$ 100 nm 
or, equivalently, that the normal membrane force is $\sim 80$ pN
\cite{spicule1}. Note that 
the boundary between spicule formation and fiber run-away
is not sensitive to the polymerisation rate, unlike the boundary 
between spicule formation and stalling. Thus, for
physiological parameters, there is only a finite range of incidence angle
for which a bundled actin fiber can form a filopodium.

\begin{figure}
\centerline
{\includegraphics[width=4.8cm]{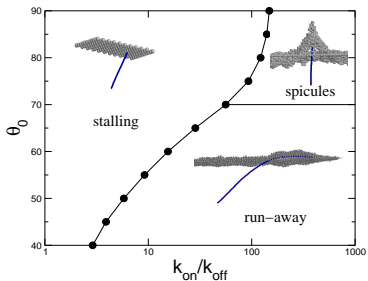}}
\caption{Phase diagram for a bundle of $n=10$ actin fibers, 
with parameters obtained by scaling those used for Fig. 2 to 
this bundle size (see text). The insets are 3dMC simulations
obtained with parameter values as in Fig~2 except the spicule for which 
$\sqrt{A}\delta=0.1$.}
\end{figure}

We know of at least two other mechanisms beyond fiber bundling
which the cell may employ to avoid excessive run-away of actin fibers: (i) capping proteins \cite{bray}, (ii) tethering of fiber tips to the membrane \cite{dickinson}.

It is also interesting to consider a series of {\it in vitro} experiments aimed at 
controlling cell motility via protein expression \cite{news_and_views}
in lights of our results. These experiments showed that lamellipodia only form when capping
proteins are present in the cell, so that they can be thought of
as a dense network of short and mostly artificially stalled fibers.
Depleting the cell of capping proteins, without decreasing the amount
of Ena/VASP proteins, results in massive growth of filopodia, 
akin to spicules. Reducing the number of Ena/VASP proteins as
well, leads on the other hand to the formation of membrane ruffles, which 
may be associated with extensive fiber run-away, 
in agreement with what we predict for actin fibers 
polymerising with cell-like conditions. All the states in our
phase diagram can thus be obtained in a cell by regulating 
the concentrations of these key proteins.
  
The validity of our treatment is not
limited to single and bundles actin fibers.
Microtubules and sickle hemoglobin fibers may have 
parameters closer to $\delta\sim 0.5$ nm and $\kappa\sim 0.1-1$~mm. If we adopt 
$d\sim 1$ $\mu$m, comparable to the size of the cell, stalling occupies a smaller region in the dynamic 
`phase diagram' , and run-away and spicule
formation a larger region. This is in agreement with 
experiments \cite{phi}.

Nonlinear membrane deformations, leading to spicule (or filopodia) formation 
cannot be described quantitatively using  Eqs. \ref{saddle_pol}-\ref{force}. To check our 
approximate analytical threshold for the onset of spicule formation,
we have performed full 3-dimensional Monte-Carlo dynamic simulations, 
in which both membrane and fiber are discretised into connected `beads'.
A time step in the dynamic evolution consists of attempting 
to move the position of all beads in both the fiber and the membrane 
via local moves (see Refs. \cite{kink-jump,nigel}
for details on the algorithm). A local move is
accepted according to the Metropolis test with discretised 
versions of Eq.~\ref{hamiltonian}.

Examples of run-away state and
spicule, found via our simulations, 
are shows as insets in Figs. 2 and 4. (Parameters are given in the legends.)
For $\theta_0=90^\circ$, we confirm that there is a crossover 
between run-away and spiculation.
For $\sqrt{A} \delta \simeq 0.2$
this occurs when the persistence length of the 
polymers exceeds $1.5-2$ $\mu$m, in good agreement with  
$1.25$ $\mu$m found using our Gaussian approximation. In few cases for lower $l_p$ values 
the fiber does not run away, 
but is `trapped' by the membrane fluctuations giving rise to a spicule.
We also find that the run-away state (see Figs. 2 and 4) 
gives rise to a bump in the membrane which may account for the
observation of ruffles \cite{news_and_views}. 
{ {Finally, if we require that a fiber does not run-away for
$10^8$ Monte-Carlo steps, 3dMC simulations predict that the crossover
to the run-away state in Fig. 1 
occurs for $\theta_c \simeq 85^{\circ}$ for $k_{\rm on}/k_{\rm off}=100$,
which is in good agreement with $\theta_c\simeq 86^{\circ}$ found via
Eqs. \ref{saddle_pol}-\ref{force}.} }

In conclusion, we have solved the dynamic equations of motion for a 
growing semi-flexible polymer, incident onto a fluctuating 
soft membrane at an angle, and found a dynamic transition between fiber run-away and
stalling. Given the kinetic polymerisation rate of 
actin fibers in the cell, we predict that
single actin fibers should bend and run away 
from the membrane, and would not contribute to cell motility. 
This is in agreement with {\it in vitro} investigations under
unphysiological conditions in which the behaviour of cells lacking
bundling and capping proteins was investigated.
Moving cells seems to have taken `precautions' to stop this 
phenomenon from happening excessively, by bundling or capping fibers
to render them stiffer. 
Our phase diagram can be generalised to the case of an actin bundle,
and we show that in this case a third regime appears,
in which the fiber deforms the membrane in a non-linear fashion and
spicules appear, which are akin to the filopodial protrusions observed
in motile cells. It is intriguing that 
physiological values of the thermodynamic parameter $\log k_{\rm on}/k_{\rm off}$ 
can be seen to lie in, or close to, the narrow range 
of values for which actin filipodia can explore all three different destinies. 
Finally, we predict that microtubules and
sickle hemoglobin fibers should be more likely to form spicules or
to run away than to stall, in agreement with observations.
 
This work was supported by NIH (NHLBI) grant HL58512 and EPSRC grant GR/S29256/01.


\begin{thebibliography}{99}
\bibitem{bray} D. Bray, {\it Cell movements: from molecules to
motility}, 2nd edition, Garland Publishing (2001); B. Alberts {\it et al.}, 
{\it Molecular biology of the cell}, Garland, New York (1994).
\bibitem{oster} C. Peskin, G. M. Odell, G. F. Oster,
{\it Biophys. J.} {\bf 65}, 316 (1993).
\bibitem{mogilner} A. Mogilner, G. Oster, {\it Biophys. J.} {\bf 71},
3030 (1996).
\bibitem{eaton} W. A. Eaton, J. Hofrichter, {\it Adv. Protein Chem.} 
{\bf 40}, 63 (1990).
\bibitem{phi} D. K. Fygenson, J. F. Marko, A. Libchaber,
{\it Phys. Rev. Lett.} {\bf 79}, 4497 (1997).
\bibitem{dickinson} R. B. Dickinson, L. Caro, D. L. Purich, 
{\it Biophys. J.} {\bf 87}, 2838 (2004).
\bibitem{mogilner_filopodia} A. Mogilner, B. Rubinsten,
{\it Biophys. J.} {\bf 89}, 782 (2005).
\bibitem{nigel} N.J. Burroughs, D. Marenduzzo, {\it J. Chem. Phys.}
{\bf 123}, 174908 (2005).
\bibitem{derenyi} I. Derenyi, F. Julicher, J. Prost,
{\it Phys. Rev. Lett.} {\bf 88}, 238101 (2002).
\bibitem{rob} D. R. Daniels, M. S. Turner, {\it J. Chem. Phys.} 
{\bf 121}, 7401 (2004).
\bibitem{note} { {However a computation of the
average membrane shape subject to the force in Fig. 3b
(done as in A. R. Evans {\em et al.}, 
{\it Phys. Rev. E} {\bf 67}, 041907 (2003))
reveals that fiber and membrane do not intersect until well after 
the run-away transition.}}
\bibitem{tomography} O. Medalia {\it et al.}, 
{\it Science} {\bf 298}, 1209 (2002).
\bibitem{filopodia} T. M. Svitkina {\it et al.}, {\it J. Cell.
Biol.} {\bf 160}, 409 (2003); I. V. Maly, G. G. Borisy, {\it Proc. Natl. 
Acad. Sci. USA} {\bf 98}, 11324 (2001).
\bibitem{spicule1} D. Raucher, M. P. Scheetz, {\it Biophys. J.}
{\bf 77}, 1992 (1999).
\bibitem{news_and_views} M. R. Mejillano {\it et al.}, 
{\it Cell} {\bf 118}, 363 (2004); D. A. Schafer, {\it Nature}
{\bf 430}, 734 (2004).  
\bibitem{kink-jump} A. Baumgartner, 
{\it Annu. Rev. Phys. Chem.} {\bf 35}, 419 (1984).
\end{thebibliography}
\end{document}